\documentclass[%
reprint,
amsmath,amssymb,
aps,
]{revtex4-2}

\usepackage{graphicx}
\usepackage{dcolumn}
\usepackage{bm}
\usepackage{url}

\newcommand{\ucd}[1]{\stackrel{\triangledown}{#1}}

\newcommand{\eV}{e^{(V)}}

\begin{document}
	
	\title{Rational Extended Thermodynamics for Non-Newtonian Fluids\\ with Finite Relaxation Time}
	
	\author{Tommaso Ruggeri}
	\email{tommaso.ruggeri@unibo.it}
	\affiliation{Department of Mathematics, University of Bologna \\ and Accademia dei Lincei}
	
	\date{\today}
	
	\begin{abstract}
		We introduce a one-dimensional, hyperbolic model for non-Newtonian fluids with finite relaxation time, derived within the framework of Rational Extended Thermodynamics (RET). Unlike classical parabolic models, our formulation preserves finite signal speeds, thermodynamic consistency, and mathematical well-posedness. The model captures viscoelastic phenomena via a nonlinear evolution of stress, converging to power-law rheology in the vanishing relaxation limit. 
        Notably, it mimics the Phan–Thien–Tanner model under steady shear, but derives from first principles, offering a predictive alternative to empirical rheology.
        % Notably, it exhibits behavior analogous to the Phan–Thien–Tanner model under steady shear, yet arises from first principles. This RET-based approach opens the door to predictive, structure-based modeling of complex fluids beyond empirical rheology.
	\end{abstract}
	
	\maketitle
	
	Non-Newtonian fluids—such as polymer solutions, slurries, blood, or paints—exhibit rheological behaviors that deviate significantly from the classical Newtonian paradigm. Shear-thinning, viscoelasticity, yield stress, and stress relaxation are common features that arise from microstructural dynamics, which classical models fail to capture.
	
	A particularly important aspect is the nonlinear dependence of the viscous stress tensor $\bm{\sigma}$ on the rate-of-deformation tensor $\mathbf{D}$, as well as the presence of \emph{relaxation effects}, in which the stress does not respond instantaneously to deformation but instead evolves over a finite time. These effects are especially relevant in viscoelastic fluids subjected to high-frequency oscillations, transient shear flows, or sharp velocity gradients.
	
	Standard approaches, such as the Navier--Stokes equations with nonlinear constitutive laws, often neglect relaxation and result in nonphysical predictions, including infinite propagation speeds. Moreover, many widely used rheological models—like power-law or PTT-type models—lack a rigorous thermodynamic foundation.

    In this Letter, we present a physically grounded alternative based on Rational Extended Thermodynamics (RET), a framework developed by M\"uller and Ruggeri~\cite{RET} that systematically incorporates higher-order moments as dynamic variables. By treating stress as an independent field, RET allows for hyperbolic balance laws that respect the second law of thermodynamics and exhibit finite signal speeds.

Building on recent advances~\cite{ViscoRuggeri}, we derive a one-dimensional, isothermal RET model for non-Newtonian fluids with finite relaxation time. The model recovers classical power-law behavior in the relaxation limit and exhibits nonlinear stress dynamics akin to the Phan–Thien–Tanner model under steady shear. The resulting system is symmetric hyperbolic, satisfies the entropy inequality, and fulfills the K-condition, ensuring well-posedness and the existence of both smooth and shock solutions.

This RET-based model offers a predictive alternative to empirical rheology by linking macroscopic behavior to a thermodynamically consistent microstructure. It is also extendable to multidimensional and thermal effects, making it suitable for industrial and biological fluid applications.

Common viscoelastic models extending the Maxwell framework include the \emph{Oldroyd-B model} (with Newtonian solvent), the \emph{Giesekus model} (accounting for anisotropic drag), and the \emph{Phan–Thien–Tanner (PTT) model}, which uses nonlinear stress functions to better capture shear-thinning behavior. The latter is expressed as  \cite{phan1977}:

	\begin{equation}
		f\bigl(\text{tr}(\boldsymbol{\sigma})\bigr)\, \boldsymbol{\sigma} + \lambda \ucd{\boldsymbol{\sigma}} = 2\mu\, \mathbf{D},
		\label{PPT}
	\end{equation}
	where $\ucd{(\cdot)}$ denotes the upper-convected derivative, and $f$ is a function of the stress trace. The PTT model is widely used to simulate polymeric and biological fluids, offering a robust empirical framework that extends linear viscoelastic models by introducing stress-dependent nonlinearities~\cite{bird1987dynamics,larson1999structure,owens2002computational}.
	
	Despite their practical success, these models exhibit three key limitations: (i) they are empirical and not derived from fundamental thermodynamic principles; (ii) they lack a quasilinear or balance-law structure, making them ill-suited for studying shocks or discontinuous solutions; and (iii) they rely on nonlocal constitutive laws that often violate objectivity. To address these issues, artificial derivatives such as the upper-convected derivative are typically introduced—an approach at odds with the philosophy of \emph{Rational Extended Thermodynamics} (RET), which formulates constitutive equations as local balance laws with production terms, rather than nonlocal closures.
	
	In continuum mechanics, fundamental laws are cast as balance equations and must be closed by constitutive relations that obey invariance principles and the second law of thermodynamics. Classical laws—such as those of Navier--Stokes, Fourier, Fick, and Darcy—are nonlocal and couple fluxes to gradients of state variables. However, Müller~\cite{mullerfourier} demonstrated that such laws generally violate objectivity, and Ruggeri~\cite{Ruggeri_Can} argued that they represent limiting cases of more fundamental \emph{hyperbolic} balance laws with local constitutive relations, valid when relaxation times vanish.
	
	For instance, the Navier--Stokes and Fourier laws emerge from RET equations in the zero-relaxation limit for shear stress, dynamic pressure, and heat flux; Fick’s law arises as the limit of momentum equations for diffusing species; and Darcy’s law can be derived by neglecting inertia in mixture momentum balances. These results support the conjecture in~\cite{Ruggeri_Can} that physical systems are fundamentally governed by hyperbolic balance laws—a view particularly compelling in relativistic and fast transient regimes.
	
	This perspective is reinforced in rarefied gas dynamics, where RET arises naturally from kinetic theory. Starting from the Boltzmann equation, one obtains an infinite hierarchy of balance equations, and appropriate moment closures—using either continuum principles (RET) or the \emph{Maximum Entropy Principle (MEP)}—lead to local, hyperbolic models with finite propagation speeds.
	
	The RET framework, initially developed for monatomic gases~\cite{RET} and later extended to polyatomic and mixture of gases~\cite{beyond,newbook}, provides a unified and thermodynamically consistent structure for continuum theories. For a comprehensive overview of MEP in the context of moment closure, see~\cite{MEP-Dreyer}.
	
	More recently, Ruggeri~\cite{ViscoRuggeri} revisited earlier efforts to incorporate \emph{viscoelasticity} into RET. The core principles of this formulation are:
	(i) New flux variables require new balance equations with \emph{local} and \emph{objective} constitutive laws;
	(ii) The resulting system must be closed in accordance with general physical requirements—namely, Galilean invariance, the entropy principle, and convexity of the entropy density;
	(iii) The final structure must be \emph{symmetric hyperbolic}, ensuring both thermodynamic consistency and mathematical well-posedness.
	For simplicity, the analysis in~\cite{ViscoRuggeri} focuses on the isothermal case and adopts Lagrangian coordinates \((\mathbf{X}, t)\), where \(\mathbf{X}\) denotes a material point in the reference configuration and \(t\) denotes time. Let \(\mathbf{x}(\mathbf{X}, t)\) be the current position of the point, and define the deformation gradient \(\mathbf{F} = \partial \mathbf{x} / \partial \mathbf{X}\). We denote by \(\mathbf{v}\) the velocity, by \(\mathbf{T}\) the first Piola–Kirchhoff stress tensor, by \(\rho^*\) the mass density in the reference configuration, and by \(\mathbf{b}\) the external body force.
		Following~\cite{ViscoRuggeri}, we consider the one-dimensional case (e.g., uniaxial deformation), where all tensorial quantities reduce to scalars.
	
	In the compressible  viscoelastic setting of~\cite{ViscoRuggeri}, the total stress was written as a sum of elastic and viscous parts, \(T(F) + \sigma\). To incorporate an equation for \(\sigma\) consistent with the RET framework, we recognize that introducing \(\sigma\) as a dynamic flux variable requires a corresponding balance law. We thus consider the following system of differential balance laws:
	\begin{align}
		\begin{split}
			& \rho^* v_t - (T(F) + \sigma)_X = \rho^* b, \\
			& F_t - v_X = 0, \\
			& \psi(F,\sigma)_t + \Omega(F,\sigma)_X = P(F,\sigma),
		\end{split}
		\label{elastvisco}
	\end{align}
	where \(\psi\), \(\Omega\), and \(P\) are constitutive functions, depending only locally on \((F, \sigma)\), due to Galilean invariance.
	
	To determine these functions, we require the any solution (classical and weak)  of system \eqref{elastvisco} satisfy the energy dissipation principle, given by the supplementary balance law:
	\begin{align}
		\begin{split}
			& \rho^* \left({v^2}/2 + e(F,\sigma) \right)_t - \left( (T(F) + \sigma)v \right)_X - \rho^* b v = \mathcal{E} \leq 0.
		\end{split}
		\label{energia}
	\end{align}
	In the isothermal case, this dissipation inequality plays the role of the entropy principle in non-isothermal processes.
	
	To restrict the constitutive set
	\[
	\mathcal{Z} \equiv \left\{ \psi(F,\sigma),\, \Omega(F,\sigma),\, P(F,\sigma),\, e(F,\sigma),\, \mathcal{E}(F,\sigma) \right\},
	\]
	we impose the general principles of RET: compatibility with the dissipation inequality~\eqref{energia} and convexity of the total energy with respect to the densities. Assuming a separable form for the internal energy,
	\begin{equation}
		e(F,\sigma) = e^{(E)}(F) + e^{(V)}(\sigma),
		\label{somma}
	\end{equation}
	it was shown in~\cite{ViscoRuggeri} that the most general form of system~\eqref{elastvisco} satisfying these requirements is:
	\begin{align}
		\begin{split}
			& \rho^* v_t - (T(F) + \sigma)_X = \rho^* b, \\
			& F_t - v_X = 0, \\
			& Z_t(\sigma) - v_X = P(F,\sigma),
		\end{split}
		\label{15}
	\end{align}
	with constitutive relations
	\begin{align*}
		& T = \rho^* e^{(E)}_F(F), \quad e^{(E)}_{FF}(F) > 0, \\
		& Z_\sigma =  {e^{(V)}_\sigma(\sigma)}/{\sigma} > 0, \\
		& \mathcal{E} = \sigma P(F,\sigma) \leq 0, \quad P(F,0) = 0.
	\end{align*}
	
	The system~\eqref{15} is symmetric hyperbolic when written in terms of the \emph{main field}
$\mathbf{u}^\prime = (v, T, \sigma)$,
as established by the general theory of hyperbolic systems of balance laws compatible with a convex entropy functional~\cite{RS}. These variables correspond to the entropy multipliers, which yield the dissipation law~\eqref{energia} upon contraction with the original system.

	In~\cite{ViscoRuggeri}, the production term was taken as \(P = -F \sigma / \mu\), so that in the limit of vanishing relaxation time (i.e., when \(Z_\sigma \to 0\)), the viscous stress reduces to the Navier–Stokes law: \(\sigma = \mu v_x\).
	
	The case of compressible fluids is recovered when \(T = -p\) and \(F = \rho^*/\rho\), with \(\rho\) denoting the current mass density.

To generalize the model to \emph{non-Newtonian} behavior—such as power-law fluids—we consider the constitutive relation
\begin{equation}\label{PowerLaw}
  \boldsymbol{\sigma} = k\,\dot{\gamma}^{\,m-1}\,\mathbf{D}, \quad
  \dot{\gamma} = 2\sqrt{\mathbf{D}:\mathbf{D}} = 2\|\mathbf{D}\|,
\end{equation}
where \(k > 0\) is the consistency coefficient and \(m > 0\) is the flow index. 
 This assumption holds for incompressible fluids or the Stokes case (zero bulk viscosity). In one-dimensional flow, only one viscosity coefficient exists, which applies here.
The main idea of this Letter is based on the observation that relation~\eqref{PowerLaw} can be inverted to express \(\mathbf{D}\) in terms of \(\boldsymbol{\sigma}\):
\begin{equation}
  \mathbf{D}
  = 2^{\frac{1}{m}-1} k^{-1/m} \|\boldsymbol{\sigma}\|^{\frac{1 - m}{m}} \boldsymbol{\sigma}.
  \label{eq:powerlaw_inverted}
\end{equation}
In the one-dimensional case, this reduces to
\begin{equation}
  v_x =  {v_X}/{F} =
  2^{\frac{1}{m}-1} k^{-1/m} |\sigma|^{\frac{1 - m}{m}} \sigma.
  \label{eq:powerlaw_1d}
\end{equation}
To ensure that system~\eqref{elastvisco} incorporates the power-law equation~\eqref{eq:powerlaw_1d} in the parabolic limit, it is sufficient to choose the following production term:
\[
  P = -F\, 2^{\frac{1}{m}-1} k^{-1/m} |\sigma|^{\frac{1 - m}{m}} \sigma,
\]
which yields the following RET-based model for non-Newtonian viscoelastic fluids:
\begin{align}
  \begin{split}
    & \rho^* v_t + (p(F) - \sigma)_X = 0, \\
    & F_t - v_X = 0, \\
    & Z_t(\sigma) - v_X = -F\, 2^{\frac{1}{m}-1} k^{-1/m} |\sigma|^{\frac{1 - m}{m}} \sigma,
  \end{split}
  \label{15c}
\end{align}
with pressure given by \(p(F) = -\rho^* e^{(E)}_F(F)\), and
\begin{equation}
  \tau(\sigma) = Z_\sigma = \rho^*  {e^{(V)}_\sigma(\sigma)}/{\sigma} > 0.
  \label{tautau}
\end{equation}

This system satisfies the energy dissipation inequality:
\[
  \left( \rho^* {v^2}/{2} + \rho^* e^{(E)}(F) + \rho^* e^{(V)}(\sigma) \right)_t
  + \left( (p(F) - \sigma)v \right)_X = \mathcal{E},
\]
with
$
  \mathcal{E} = -F\, 2^{\frac{1}{m}-1} k^{-1/m} |\sigma|^{\frac{1}{m} + 1} \leq 0
$,
and main field \(\mathbf{u}^\prime = (v, -p, \sigma)\). 
We need to observe that while the power law \eqref{PowerLaw} together with equations \eqref{elastvisco}$_{1,2}$ is not a quasi-linear system, the system \eqref{15c} is a quasi-linear systems as all the first order derivative are linear thanks to the idea to consider the inverse function of power-law \eqref{eq:powerlaw_1d}.

The last equation in~\eqref{15c} may be rewritten  for smooth solutions as 
\begin{equation}
  \tau(\sigma)\, \sigma_t - v_X
  = -F\, 2^{\frac{1}{m}-1} k^{-1/m} |\sigma|^{\frac{1 - m}{m}} \sigma.
  \label{bellissima}
\end{equation}
Note that $\tau$ has units of time per viscosity (i.e., $1/\mathrm{Pa}$). As $\tau \to 0$, Eq.~\eqref{bellissima} recovers the algebraic power-law relation~\eqref{eq:powerlaw_1d}, generalizing RET to non-Newtonian fluids while preserving hyperbolicity and thermodynamic consistency.

% We have to notice that the dimension of $\tau$ is not a time but a time divided a viscosity, i.e. $1/Pa$.
% As \(\tau \to 0\), Eq.~\eqref{bellissima} reduces to the algebraic power-law relation~\eqref{eq:powerlaw_1d}. The model thus generalizes RET to non-Newtonian behavior while preserving hyperbolicity and thermodynamic consistency.

	Moreover, this structure is not restricted to power-law fluids. Any invertible relation \(\boldsymbol{\sigma}\equiv \boldsymbol{g}(\mathbf{D})\), with inverse \(\mathbf{D} \equiv \boldsymbol{g}^{-1}(\boldsymbol{\sigma})\), can be incorporated (in 1D and under isothermal conditions) by setting \(P = F\, g^{-1}(\sigma)\), provided that \(\sigma\, g^{-1}(\sigma) \leq 0\) to ensure energy dissipation.
	
	As a special case, taking a quadratic viscous energy:
	\[
	e^{(V)}(\sigma) = \frac{\tau_0}{2 \rho^*} \sigma^2,
	\]
	yields a constant   \(\tau(\sigma) = \tau_0\), and the evolution equation~\eqref{bellissima} becomes:
	\begin{equation}
		\tau_0 \sigma_t - v_X = -F\, 2^{\frac{1}{m}-1} k^{-1/m} |\sigma|^{\frac{1 - m}{m}} \sigma.
		\label{bella2}
	\end{equation}
	
	This remains a nonlinear equation in $\sigma$, resembling the PTT model \eqref{PPT} in the one-dimensional case, but with a standard time derivative rather than an upper-convected one, and a particular choice of the stress function $f(\text{tr}(\bm{\sigma}))$.
	
	We emphasize that in the general case of \eqref{bellissima}, the relaxation time is a nonlinear function of $\sigma$ and is determined by the viscous energy density $\eV(\sigma)$.
	
	To gain insight into the mathematical behavior of equation \eqref{bella2}, we examine two representative cases:

	\paragraph{Case 1: Constant Velocity ($v_X = 0$)}
	
	Assuming that at time $t=0$ the viscous stress is $\sigma_0$ and the deformation gradient is $F = F_0 = 1$, equation \eqref{bella2} (for $\sigma > 0$) reduces to:
	\begin{align*}
    & \sigma_{\bar{t}}= - a(m,k) \, \sigma ^{\frac{1}{m}},\qquad \sigma(0)=\sigma_0,\\
    & a(m,k)=   {2^{\frac{1}{m}-1} k ^{-\frac{1}{m}}}, \quad \bar{t} = {t}/{\tau_0} .
\end{align*}
	The solution behavior depends critically on the parameter $m$:
	\begin{itemize}
		\item \textbf{Shear-thinning ($m < 1$):}
		\begin{equation}
			\sigma(\bar{t}) = \left[\sigma_0^{\frac{m-1}{m}} + a(m,k)\, \frac{1-m}{m}\, \bar{t} \right]^{-\frac{m}{1-m}},
		\end{equation}
		which decays to zero algebraically as $\bar{t} \to \infty$.
		
		\item \textbf{Maxwell case ($m = 1$):}
		\begin{equation}
			\sigma(\bar{t}) = \sigma_0\, e^{-k \bar{t}},
		\end{equation}
		representing standard exponential relaxation.
		
		\item \textbf{Shear-thickening ($m > 1$):}
		\begin{align}
			\sigma(\bar{t}) &=
			\begin{cases}
				\left[\sigma_0^{\frac{m-1}{m}} - a(m,k)\, \frac{m-1}{m}\, \bar{t} \right]^{\frac{m}{m-1}}, & \text{for } \bar{t} < \bar{t}_c, \\
				0, & \text{for } \bar{t} \geq \bar{t}_c,
			\end{cases} \\
			\bar{t}_c &= \frac{m}{m-1}\, \frac{\sigma_0^{\frac{m-1}{m}}}{a(m,k)}.
		\end{align}
		In this case, $\sigma$ decays to zero in a finite time.
	\end{itemize}
	\begin{figure}[ht]
		\centering
		\includegraphics[width=0.7\linewidth]{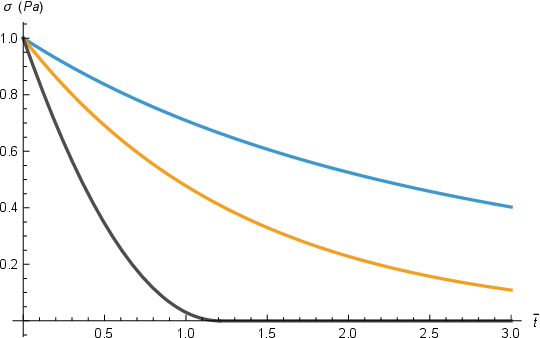}
		\caption{Decay of viscous stress versus  $\bar{t}$, for different $m$. From up to down: $m=0.7, m=1, m=2$ respectively.}
		\label{fig:enter-label}
	\end{figure}
	\begin{figure}[htbp]
		\centering
		\includegraphics[width=0.7\linewidth]{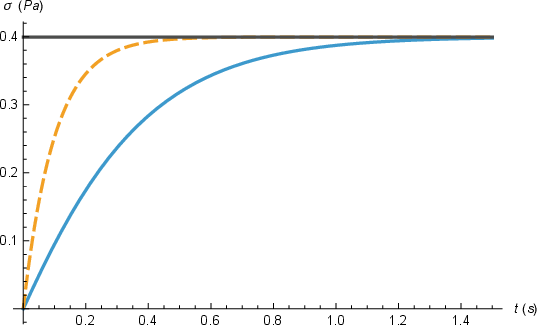}
		\caption{Stress response over time. The values are $v_x^0=0.1, m=0.7, \tau_0=\tau_1=0.1$.
			\textbf{(--)} Present model, 
			\textbf{(- -)} Linear relaxation. } 
		\label{mmin}
	\end{figure}
	These distinct regimes are governed solely by $m$, while $k$ and $m$ are assumed to be independent. For illustrative purposes, we set $k = 10 e^{-2m}$ and $\sigma_0 = 1$. The stress evolution for different $m$ is shown in Fig.~\ref{fig:enter-label}.
	
	\paragraph{Case 2: Constant Velocity Gradient ($v_x = v_x^0$)}
	Here, the deformation gradient evolves as:
	\begin{equation}
		\dot{F} = v_x^0 F \quad \Rightarrow \quad F(t) = F_0 e^{v_x^0 t}.
	\end{equation}
	In this scenario, equation \eqref{bella2} becomes:
	\begin{equation}\label{bella3}
		\tau_0\,  {\sigma}_t = F_0\, e^{v_x^0 t} \left(v_x^0 - a(m,k)\, \sigma^{\frac{1}{m}}\right).
	\end{equation}
	An analytical solution of \eqref{bella3} involves hypergeometric function. Considering the most physical case in which $\sigma_0=\sigma(0)=0$, we observe that the relaxation becomes faster with increasing $v_x^0$.  It is easy to prove rigorously that the steady state is achieved when the right-hand side vanishes:
	\begin{equation}\label{pl}
		\sigma^\infty = \lim_{t \to \infty} \sigma(t) = \left(\frac{v_x^0}{a(m,k)}\right)^m,
	\end{equation}
	which corresponds to the classical power-law behavior. In  reality it is possible to prove that the convergence is \emph{superexponential}.
	Figure~\ref{mmin} illustrates the stress evolution for $m=0.7$ (shear-thinning), 
   % while Fig.~\ref{mmag} shows the $m=1.2$ case (shear-thickening). 
    We compare our model with a linear Maxwell-type relaxation $\tau_1\,  {\sigma}_t = \sigma^\infty - \sigma$. For $\sigma_0 = 0$, this leads to exponential behavior: $\sigma(t) = \sigma^\infty (1 - e^{-t/\tau_1})$. While both models converge to the same steady state, the relaxation dynamics differ significantly due to the nonlinear nature of equation \eqref{bella3}.
	
	% In the shear-thinning case, the present model reaches the asymptotic state more slowly than the linear model; conversely, for shear-thickening, it converges faster. These differences highlight the nontrivial effects of nonlinearity, though the precise behavior depends sensitively on model parameters.
	% 	\begin{figure}[htbp]
	% 	\centering
	% 	\includegraphics[width=0.8\linewidth]{grafico_sigma_vs_t2.eps}
	% 	\caption{Stress response over time. The values are $v_x^0=0.1, m=1.2, \tau_0=\tau_1=0.1$.
	% 		The curves represent: 
	% 		\textbf{(--)} Present model (blue), 
	% 		\textbf{(- -)} Linear relaxation (orange), and 
	% 		\textbf{(gray solid)} power-law asymptote. }
	% 	\label{mmag}
	% \end{figure}

	\paragraph{Conclusion}
	We have demonstrated that the proposed viscoelastic model—derived from Rational Extended Thermodynamics (RET)—offers a physically consistent and mathematically robust framework also for describing relaxation in non-Newtonian fluids, particularly in one-dimensional, isothermal settings. In contrast to phenomenological models, our system is built from first principles and leads to a quasi-linear, symmetric hyperbolic system of balance laws. This structure ensures compliance with fundamental thermodynamic constraints and allows for the emergence of weak solutions and shock phenomena.
		For the special case of quadratic viscous energy, our model exhibits structural similarities with the Phan–Thien–Tanner (PTT) model. Given the PTT model’s success in modeling dilute polymer solutions and applications in industrial and microfluidic contexts, we expect the present model to perform similarly well in capturing experimental behavior. This conjecture will be investigated in future studies.


\begin{thebibliography}{99}
\bibitem{RET} I. M\"uller and T. Ruggeri, \textit{Rational Extended Thermodynamics}, 2nd ed. (Springer, New York, 1998).
		
		\bibitem{ViscoRuggeri} T. Ruggeri, Int. J. Non-Linear Mech. \textbf{160}, 104658 (2024).
		
				
		\bibitem{phan1977} N. Phan-Thien and R. I. Tanner, J. Non-Newtonian Fluid Mech. \textbf{2}, 353 (1977).
		
				
				
		\bibitem{bird1987dynamics} R. B. Bird, R. C. Armstrong, and O. Hassager, \textit{Dynamics of Polymeric Liquids, Vol. 1: Fluid Mechanics} (Wiley, 1987).
		
		\bibitem{larson1999structure} R. G. Larson, \textit{The Structure and Rheology of Complex Fluids} (Oxford Univ. Press, 1999).
		
		\bibitem{owens2002computational} R. G. Owens and T. N. Phillips, \textit{Computational Rheology} (Imperial College Press, 2002).
		
		\bibitem{mullerfourier} I. M\"uller, Arch. Ration. Mech. Anal. \textbf{45}, 241 (1972).
		
		\bibitem{Ruggeri_Can} T. Ruggeri, Quart. Appl. Math. \textbf{70}, 597 (2012).
		
				
		\bibitem{beyond} T. Ruggeri and M. Sugiyama, \textit{Rational Extended Thermodynamics beyond the Monatomic Gas} (Springer, Cham, 2015).
		
		\bibitem{newbook} T. Ruggeri and M. Sugiyama, \textit{Classical and Relativistic Rational Extended Thermodynamics of Gases} (Springer, Cham, 2021).
		
		\bibitem{MEP-Dreyer} T. Arima and T. Ruggeri, arXiv:2412.18340 (to appear in \textit{Advances in Continuum Physics: In Memoriam of Wolfgang Dreyer}, Springer, 2025).
		
			
		\bibitem{RS} T. Ruggeri and A. Strumia, Ann. Inst. H. Poincaré Sect. A \textbf{34}, 65 (1981).
		
		% \bibitem{giusti} A. Giusti, A. Mentrelli, and T. Ruggeri, Int. J. Non-Linear Mech. \textbf{161}, 104685 (2024).
		
		% \bibitem{amabili} M. Amabili, T. Arima, and T. Ruggeri, J. Mech. Phys. Solids \textbf{196}, 106033 (2025).
		
				
	%	\bibitem{NeRuggeri} T. Ruggeri, \textit{Introduction to the Thermomechanics of Continua and Hyperbolic Systems} (Springer, Cham, 2024).
		
		
	


\end{thebibliography}
\end{document}